\input phyzzx
\overfullrule=0pt
\def\Dslash{D\kern-0.15em\raise0.17ex\llap{/}\kern0.15em\relax}
\def\tr{\mathop{\rm tr}}
\def\Re{\mathop{\rm Re}}
\def\Im{\mathop{\rm Im}}
\def\Det{\mathop{\rm Det}}
\REF\HAS{%
P. Hasenfratz,
Nucl.\ Phys.\ (Proc.\ Suppl.) {\bf 63} (1998), 53;
Nucl.\ Phys.\ {\bf B525} (1998), 401.}
\REF\HASS{%
P. Hasenfratz, V. Laliena and F. Niedermayer,
Phys.\ Lett.\ {\bf B427} (1998), 125.}
\REF\NEU{%
H. Neuberger,
Phys.\ Lett.\ {\bf B417} (1998), 141; {\bf B427} (1998), 353.}
\REF\GIN{%
P. H. Ginsparg and K. G. Wilson,
Phys.\ Rev.\ {\bf D25} (1982), 2649.}
\REF\LUS{%
M. L\"uscher,
Phys.\ Lett.\ {\bf B428} (1998), 342.}
\REF\KIK{%
Y. Kikukawa and A. Yamada,
Phys.\ Lett.\ {\bf B448} (1999), 265.}
\REF\LUSC{%
M. L\"uscher,
hep-lat/9811032.}
\REF\NAR{%
R. Narayanan,
Phys.\ Rev.\ {\bf D58} (1998), 097501.}
\REF\NIE{%
F. Niedermayer,
talk given at the International Symposium on Lattice Field Theory,
Boulder 1998, hep-lat/9810026.}
\REF\KIKU{%
Y. Kikukawa and A. Yamada,
hep-lat/9808026.}
\REF\NEUB{%
H. Neuberger,
hep-lat/9802033.}
\REF\NARA{%
R. Narayanan and H. Neuberger,
Nucl.\ Phys.\ {\bf B412} (1994), 574; {\bf B443} (1995), 305.}
\REF\RAN{%
S. Randjbar-Daemi and J. Strathdee,
Phys.\ Lett.\ {\bf B402} (1997), 134.}
\REF\LUSCH{%
M. L\"uscher,
Nucl.\ Phys.\ {\bf B538} (1999), 515.}
\REF\SUZ{%
H. Suzuki,
Phys.\ Rev.\ {\bf D55} (1997), 2994.}
\REF\OKU{%
K. Okuyama and H. Suzuki,
Phys.\ Rev.\ {\bf D56} (1997), 6829.}
\REF\BAN{%
H. Banerjee, R. Banerjee and P. Mitra,
Z. Phys.\ {\bf C32} (1986), 445.\nextline
See also, K. Fujikawa,
talk given at Nato Advanced Research Workshop ``Super Field
Theories,'' Burnaby 1986, Hiroshima Preprint RRK~86-32.}
\REF\ADA{%
D. H. Adams,
hep-lat/9812003.}
\REF\SUZU{%
H. Suzuki,
hep-th/9812019.}
\REF\FUJ{%
K. Fujikawa,
hep-th/9811235.}
\REF\NIEM{%
A. J. Niemi and G. W. Semenoff,
Phys.\ Rev.\ Lett.\ {\bf 55} (1985), 927; {\bf 55} (1985), 2627 (E);
Nucl.\ Phys.\ {\bf B269} (1986), 131.}
\REF\BAL{%
R. D. Ball and H. Osborn,
Phys.\ Lett.\ {\bf 165B} (1985), 410.\nextline
R. D. Ball, Phys.\ Rep.\ {\bf 182} (1989), 1, and references therein.}
\REF\ALV{%
L. Alvarez-Gaum\'e, S. Della Pietra and V. Della Pietra,
Phys.\ Lett.\ {\bf B166} (1986), 177; Commun.\ Math.\ Phys.\ {\bf 109}
(1987), 691.\nextline
L. Alvarez-Gaum\'e,
``An introduction to anomalies,'' Lectures given at the International
School on Mathematical Physics, Erice 1985, {\it in\/} ``Fundamental
problems of gauge field theory,'' ed.\ G. Velo and A. S. Wightman
(Plenum Press, New York, 1986).}
\REF\BAR{%
W. A. Bardeen and B. Zumino,
Nucl.\ Phys.\ {\bf B244} (1984), 421.}
\REF\ALVA{%
L. Alvarez-Gaum\'e and P. Ginsparg,
Ann.\ of Phys.\ {\bf 161} (1985), 423.}
\REF\LEU{%
H. Leutwyler,
Phys.\ Lett.\ {\bf B152} (1985), 78.}
\Pubnum={IU-MSTP/32; hep-lat/9901012}
\date={January 1999}
\titlepage
\title{Gauge Invariant Effective Action in Abelian Chiral Gauge
Theory on the Lattice}
\author{%
Hiroshi Suzuki\foot{E-mail: hsuzuki@mito.ipc.ibaraki.ac.jp}}
\address{%
Department of Physics, Ibaraki University, Mito 310-0056, Japan}
\abstract{%
L\"uscher's recent formulation of Abelian chiral gauge theories on
the lattice, in the vacuum (or perturbative) sector in infinite
volume, is reinterpreted in terms of the lattice covariant
regularization. The gauge invariance of the effective action and the
integrability of the gauge current in anomaly-free cases become
transparent. The real part of the effective action is simply
one-half that of the Dirac fermion and, when the Dirac operator
behaves properly in the continuum limit, the imaginary part in this
limit reproduces the $\eta$-invariant.}
\endpage
We have gained a new perspective on chiral symmetries on the lattice,
after the recent discovery~[\HAS,\HASS,\NEU] of gauge covariant Dirac
operators which satisfy the Ginsparg-Wilson relation~[\GIN]. When the
Dirac operator is applied to lattice QCD, in which the chiral
symmetries are global, the action can be made invariant under all the
flavor~$U(N_f)$ axial symmetries. Quantum mechanically, on the other
hand, the flavor-singlet $U(1)$~symmetry suffers from the axial
anomaly, due to the non-trivial Jacobian factor~[\LUS,\KIK]. The
desired breaking pattern of the Ward-Takahashi identities in vectorial
gauge theories is thus restored. In this context, one can even
formulate the (analytic) index theorem with finite lattice
spacing~[\HASS,\LUS].

Quite recently, L\"uscher~[\LUSC] formulated (Abelian) chiral gauge
theories on the lattice, on the basis of the chirality separation with
respect to the Ginsparg-Wilson chiral matrix~[\NAR,\NIE,\KIKU].
He proved there exists a gauge invariant effective action of Weyl
fermions, when (and only when) the anomaly cancellation condition
{\it in the continuum field theory\/} is fulfilled.
(Neuberger~[\NEUB] made a similar observation in the overlap
formalism~[\NARA,\RAN] for a particular kind of gauge field
configuration.) The crucial ingredient in L\"uscher's proof is the
complete clarification of the structure of the axial anomaly with
finite lattice spacing~[\LUSCH].

In the present note, we give a reinterpretation of L\"uscher's
formulation in~Ref.~[\LUSC] in terms of the ``lattice covariant
regularization'' proposed in the past by the present author and a
collaborator~[\SUZ,\OKU]. This reinterpretation is possible at least
in the vacuum sector (implying that the Dirac operator has no zero
modes and the inverse of the Dirac operator exists) in infinite
lattice volume (for which the results of~Ref.~[\LUSCH] can be used
straightforwardly). Although the most interesting part
of~Ref.~[\LUSC] is the analysis of the topological sector in a finite
lattice volume, we will clarify some of properties of the formulation
in this simpler situation, by giving a one parameter integral
representation of L\"uscher's effective action.

Now, in the scheme of~Refs.~[\SUZ] and~[\OKU], the primarily-defined
quantity is the variation of the effective action with respect to the
gauge field, ``the gauge current''. The effective action is regarded
as the secondary object, which is deduced from the gauge current. The
basic idea is that using the {\it vectorial\/} gauge covariant Dirac
propagator (without species doublers) with an appropriate chirality
projection, one may always preserve the gauge covariance of the gauge
current of the Weyl fermion. In this way, the gauge symmetry is
maximally preserved even in anomalous cases, and the gauge invariance
is automatically restored (in the continuum limit) when the gauge
representation is anomaly-free.

According to this scheme, we temporarily identify a variation of the
effective action~$W$ with (see, Eq.~(9) of~Ref.~[\OKU])
$$
   \delta W\sim\Tr\delta D{\cal P}_{\rm H}D^{-1}.
\eqn\one
$$

We use a gauge covariant Dirac operator~$D$ that is assumed to
satisfy the Ginsparg-Wilson relation~[\GIN],
$$
   \gamma_5D+D\gamma_5=aD\gamma_5D,
\eqn\two
$$
with the lattice spacing~$a$. We will also assume the Hermitian
conjugate of the Dirac operator
satisfies~$D^\dagger=\gamma_5D\gamma_5$. In~Eq.~\one, ${\rm H}=\pm$
denotes the chirality of the Weyl fermion, and the projection
operator in~Eq.~\one\ is defined with respect to the modified chiral
matrix~[\NAR,\NIE,\KIKU],
$$
   {\mit\Gamma}_5=\gamma_5(1-aD)=\gamma_5-a{\cal D},
   \qquad{\cal P}_\pm={1\over2}(1\pm{\mit\Gamma}_5).
\eqn\three
$$
Here ${\cal P}_\pm$ is in fact the projection operator, because
${\mit\Gamma}_5^2=1$ due to the relation~\two. (${\mit\Gamma}_5$~is
Hermitian: ${\mit\Gamma}_5^\dagger={\mit\Gamma}_5$.) For convenience,
we introduce the Hermitian operator ${\cal D}=\gamma_5D$
and~${\cal D}^{-1}=D^{-1}\gamma_5$. The following relations, which are
the consequence of~Eq.~\two, are also useful:
$$
   {\mit\Gamma}_5{\cal D}^{-1}=-{\cal D}^{-1}{\mit\Gamma}_5-a
   =-{\cal D}^{-1}\gamma_5,
   \qquad\{{\mit\Gamma}_5,\delta{\cal D}\}=0.
\eqn\four
$$

We restrict ourselves to the case of the Abelian gauge group. In this
case, and if the lattice volume is infinite, one may always
associate~[\LUSCH] the gauge potential~$A_\mu$ with the link variable
through
$$
   U_\mu(x,t)=\exp[iatTA_\mu(x)],
\eqn\five
$$
where $T$~denotes the generator of the Abelian gauge group,
$T_{\alpha\beta}=e_\alpha\delta_{\alpha,\beta}$. Here $e_\alpha$~is
the $U(1)$~charge of the flavor~$\alpha$. In~Eq.~\five, we have
introduced the ``gauge coupling parameter''~$t$. We denote the
dependence on~$t$ in five-dimensional notation
as~$A_\mu(x,t)=tA_\mu(x)$ (where the original link variable and the
gauge potential are given by the value at~$t=1$), but the
argument~$t$ will often be omitted when there is no danger of
confusion.

In the Abelian case, we may differentiate the effective action 
with respect to the parameter~$t$ and then integrate it over this
parameter. The identification~\one\ then motivates the following
definition of the effective action~$W'$:
$$
\eqalign{
   W'&=\int_0^1dt\,\partial_tW'
\cr
   &\equiv\int_0^1dt\,\Tr\partial_tD{\cal P}_{\rm H}D^{-1}
   =\int_0^1dt\,\Tr\partial_t{\cal D}{\cal P}_{\rm H}{\cal D}^{-1}.
\cr
}
\eqn\six
$$
In the continuum field theory, the corresponding definition of the
effective action of the Weyl fermion is known to have interesting
properties~[\BAN].

The property of the would-be effective action~\six\ under the gauge
transformation is the central point of our discussion. (The following
analysis is analogous to that in~Ref.~[\OKU] with the Wilson
propagator.) We first split the functional~\six\ into real and
imaginary parts. By noting that the operators $\partial_t{\cal D}$,
${\cal P}_{\rm H}$ and~${\cal D}^{-1}$ are all Hermitian, we find
{}from Eq.~\four,
$$
   W'^*
   =\int_0^1dt\,\Tr\partial_t{\cal D}\widetilde{\cal P}_{\rm H}
   {\cal D}^{-1}
   =\int_0^1dt\,\Tr\partial_tD\widetilde{\cal P}_{\rm H}D^{-1},
\eqn\seven
$$
with $\widetilde{\cal P}_\pm={\cal P}_\mp$. The real part is
therefore given by
$$
   \Re W'={1\over2}\int_0^1dt\,\Tr\partial_tDD^{-1}
   ={1\over2}\ln\Det D.
\eqn\eight
$$
This is simply one-half the effective action of the Dirac fermion in
vectorial gauge theories. Eq.~\eight\ is manifestly gauge invariant,
because the Dirac operator is gauge covariant.

The imaginary part, on the other hand, is given by
$$
   i\Im W'={\epsilon_{\rm H}\over2}
   \int_0^1dt\,\Tr\partial_t{\cal D}{\mit\Gamma}_5{\cal D}^{-1}
   =-{\epsilon_{\rm H}\over2}\int_0^1dt\,
   \Tr\gamma_5\partial_t{\cal D}{\cal D}^{-1},
\eqn\nine
$$
where~$\epsilon_\pm=\pm1$. Let us quickly verify that the imaginary
part vanishes identically when the representation is ``vector-like,''
i.e., when there exists a unitary matrix~$u$ such that\foot{%
For example, when the $U(1)$ charges come in pairs of opposite sign,
the unitary matrix
$u_{\alpha\beta}=\delta_{\alpha,\beta-1}-\delta_{\alpha,\beta+1}$
is sufficient.} $uTu^\dagger=-T$. We assume the Dirac operator~$D$
transforms in the same way as the conventional lattice covariant
derivative under charge conjugation~[\LUSC]. This implies, in the
functional notation, $CuDu^\dagger C^{-1}=D^T$, with the charge
conjugation matrix~$C$, $C\gamma^\mu C^{-1}=-\gamma^{\mu T}$ and
$C\gamma_5C^{-1}=\gamma_5^T$. Consequently, we have
$Cu\partial_t{\cal D}{\mit\Gamma}_5{\cal D}^{-1}u^\dagger C^{-1}=
\gamma_5^T\partial_t{\cal D}^T{\mit\Gamma}_5^T{\cal D}^{-1T}
\gamma_5^T=
(\gamma_5{\cal D}^{-1}{\mit\Gamma}_5\partial_t{\cal D}\gamma_5)^T$
and
$$
\eqalign{
   \Tr\partial_t{\cal D}{\mit\Gamma}_5{\cal D}^{-1}
   &=\Tr Cu\partial_t{\cal D}{\mit\Gamma}_5{\cal D}^{-1}
   u^\dagger C^{-1}
\cr
   &=\Tr{\cal D}^{-1}{\mit\Gamma}_5\partial_t{\cal D}
\cr
   &=-\Tr\partial_t{\cal D}{\mit\Gamma}_5{\cal D}^{-1}=0.
\cr
}
\eqn\addone
$$
It is thus seen that $\Im W'=0$ for vector-like cases. This is
certainly the desired property, because in this case it is possible
to arrange the fermions so that the theory is left-right symmetric.

In general, the imaginary part~\nine\ neither vanishes nor is gauge
invariant. Its variation under the gauge transformation can be
determined by noting that the infinitesimal gauge transformation on a
gauge covariant object is represented by the commutator as
$\delta_\lambda{\cal D}=-it[\lambda T,{\cal D}]$. We have
$$
\eqalign{
   i\delta_\lambda\Im W'
   &=-i{\epsilon_{\rm H}\over2}\int_0^1dt\,
   \Tr\Bigl\{\partial_t(t[\lambda T,{\cal D}]){\mit\Gamma}_5
   {\cal D}^{-1}
   +t\partial_t{\cal D}[\lambda T,{\mit\Gamma}_5{\cal D}^{-1}]\Bigr\}
\cr
   &=i\epsilon_{\rm H}\int_0^1dt\,
   \Tr\lambda T\gamma_5\Bigl(1-{1\over2}aD\Bigr)
   \equiv i\epsilon_{\rm H}\int_0^1dt\,
   a^4\sum_x\lambda(x){\cal A}(x,t).
\cr
}
\eqn\ten
$$
{}From the first line to the second line, use of relation~\four\
has been made. In the last expression, we have defined the covariant
gauge anomaly as
$$
\eqalign{
   {\cal A}(x)
   &=\tr T\gamma_5\Bigl[1-{1\over2}aD(x)\Bigr]\delta(x,x)
\cr
   &\buildrel{a\to0}\over\rightarrow
   {1\over32\pi^2}\sum_\alpha e_\alpha^3\,
   \varepsilon_{\mu\nu\rho\sigma}F_{\mu\nu}F_{\rho\sigma}(x),
\cr
}
\eqn\eleven
$$
where the field strength has been defined by
$F_{\mu\nu}(x)=\partial_\mu A_\nu(x)-\partial_\nu A_\mu(x)$. The
anomaly in the continuum limit was perturbatively computed~[\KIK] by
using the overlap-Dirac operator introduced in~Ref.~[\NEU] (see also
Refs.~[\ADA] and~[\SUZU]) and we have used the result in~Eq.~\eleven.
One can even prove~[\FUJ] that the expression in~Eq.~\eleven\ is
insensitive to the choice of the Dirac operator, as long as this
operator behaves properly in the continuum limit. Substituting the
covariant anomaly~\eleven\ into~Eq.~\ten, we have a consistent
(Abelian) gauge anomaly (because~$\int_0^1dt\,t^2=1/3$), as should
be the case. In continuum field theory, the prescription~\six\
provides a general recipe to produce the consistent anomaly from the
covariant anomaly~[\BAN].

Not only about its non-invariance under the gauge transformation, we
can see more directly the validity of the prescription~\nine\ in the
continuum limit. The argument proceeds as follows:
$$
\eqalign{
   i\Im W'&=-{\epsilon_{\rm H}\over2}\int_0^1dt\,
   \Tr\gamma_5\partial_tDD^{-1}
\cr
   &=-\lim_{\Lambda\to\infty}{\epsilon_{\rm H}\over2}\int_0^1dt\,
   \Tr\gamma_5\partial_tDD^{-1}f({\cal D}^2/\Lambda^2)
\cr
   &\buildrel{a\to0}\over\rightarrow
   -\lim_{\Lambda\to\infty}{\epsilon_{\rm H}\over2}\int_0^1dt\,
   \Tr\gamma_5\partial_t\Dslash\Dslash^{-1}f(\Dslash^2/\Lambda^2)
\cr
   &=-i\epsilon_{\rm H}\pi\eta(0)
   +{i\epsilon_{\rm H}\over3\cdot8\pi^2}\sum_\alpha e_\alpha^3
   \int_{M^4\times R}A\,dA\,dA.
\cr
}
\eqn\twelve
$$
In the first step, we have introduced the regulator~$f(x)$, which
rapidly goes to zero as $x$~increases and satisfies~$f(0)=1$. In the
second step, we have exchanged the two limits~$a\to0$
and~$\Lambda\to\infty$ by assuming the lattice integrals without the
regulator~$f(x)$ are finite in the $a\to0$~limit. Since the
corresponding expression in the continuum field theory is UV finite,
once the gauge covariance is imposed, this is a reasonable
assumption. Next, we have assumed the Dirac operator~$D$ is free of
species doubling and thus that it diverges rapidly~$\sim1/a$ in the
momentum region corresponding to species doublers. These momentum
regions do not contribute in the $a\to0$~limit, because of the
existence of~$f(x)$:
$f(1/a^2\Lambda^2)\buildrel{a\to0}\over\rightarrow0$. In the physical
momentum region, we have assumed
$$
   D(x)\buildrel{a\to0}\over\rightarrow ic\Dslash(x),
   \qquad\hbox{$c$: real constant},
\eqn\thirteen
$$
where $\Dslash(x)$ is the covariant derivative in continuum field
theory, $\Dslash(x)=\gamma^\mu(\partial_\mu+iTA_\mu)$.
(Eq.~\thirteen\ is consistent with the assumed Hermiticity.) Our
manipulation is quite similar to that in the general analysis of
the continuum limit of the chiral Jacobian~[\FUJ]. A detailed
account of its justification can be found in Ref.~[\FUJ].

In the final step of Eq.~\twelve, we have appealed to a well-known
result in continuum field theory~[\NIEM--\ALV]: The imaginary
part of the effective action is given by the Atiyah-Patodi-Singer
$\eta$-invariant (the spectral asymmetry)~$\eta(0)$ plus the
Chern-Simons 5-form. (See, for example, the first reference
in~Ref.~[\ALV] for a heuristic proof with~$f(x)=(1+x)^{-1/2}$. Note
that the imaginary part is independent of the choice of~$f(x)$).
Equation~\twelve\ in fact reproduces the consistent gauge anomaly in
the continuum limit, Eq.~\ten\ with Eq.~\eleven.\foot{%
The $\eta$-invariant is defined
by~$\eta(0)\equiv\sum_n{\rm sign}\,\lambda_n$ from the
eigenvalue of the five-dimensional Hermitian
operator~$H=i\gamma_5\partial/\partial x^5+\nobreak\Dslash$.
The boundary condition for the gauge field is specified as
$A_\mu(x,x^5=\nobreak\infty)=A_\mu(x)$ and
$A_\mu(x,x^5=-\infty)=0$, and $A_5(x,x^5)=0$.  The four-dimensional
gauge transformation at the plane~$x^5=\infty$ can be regarded as a
five-dimensional gauge transformation that is independent of~$x^5$.
Under such a gauge transformation, the eigenvalue~$\lambda_n$ is
clearly gauge invariant, and thus is the $\eta$-invariant. On the
other hand, the Chern-Simons 5-form in~Eq.~\twelve\ is not invariant,
and it reproduces the consistent gauge anomaly~\ten\
with~Eq.~\eleven\ on the boundary~$x^5=\infty$. The gauge-invariant
information of the imaginary part is therefore carried by the
$\eta$-invariant~[\ALV].}

We have seen that the prescription~\six\ is satisfactory in the sense
that the real part is always gauge invariant and, in the continuum
limit, the imaginary part reproduces the correct result of the
continuum field theory. In particular, the gauge invariance of the
imaginary part is automatically restored in the continuum limit when
the anomaly is canceled, i.e., when~$\sum_\alpha e_\alpha^3=0$.
However, at this stage, we cannot say anything regarding the gauge
invariant property of the imaginary part~\nine\ with {\it finite}
lattice spacing: This was the main reason that the formulation
of~Ref.~[\OKU] could not be pursued.

Now, L\"uscher has given a remarkable proof~[\LUSCH] that the gauge
anomaly~\eleven\ with {\it finite\/} lattice spacing has the
following structure:\foot{%
The author is grateful to Professor~T.~Inami for an informative
discussion and for pointing out the importance of this proof in our
approach, prior to the publication of~Ref.~[\LUSC].}
$$
   {\cal A}(x)
   ={1\over32\pi^2}\sum_\alpha e_\alpha^3\,
   \varepsilon_{\mu\nu\rho\sigma}
   F_{\mu\nu}(x)F_{\rho\sigma}(x+a_\mu+a_\nu)
   +\partial_\mu^*\overline k_\mu(x).
\eqn\fourteen
$$
(We have used the coefficient in~Eq.~\eleven.) In this expression,
$\partial_\mu^*$ is the backward difference operator (see
Ref.~[\LUSCH]) and $\overline k_\mu(x)$ is a {\it gauge invariant\/}
current that depends locally on the gauge potential. The proof
in~Ref.~[\LUSCH] also gives an explicit method to
construct~$\overline k_\mu(x)$. Although the
current~$\overline k_\mu(x)$ is not unique, we can fix (partially)
its form by requiring it to transform like the axial vector current
under the lattice transformation~[\LUSCH,\LUSC].

Once having observed~Eq.~\fourteen, we may improve the effective
action~\six\ as $W=W'+\overline K$ where
$$
   \overline K=i\epsilon_{\rm H}\int_0^1dt
   \,a^4\sum_xA_\mu(x)\overline k_\mu(x,t).
\eqn\fifteen
$$
Since the gauge transformation of the gauge potential is given by
$\delta_\lambda A_\mu(x)=\partial_\mu\lambda(x)$, the gauge variation
of~Eq.~\fifteen\ is given by
$$
   \delta_\lambda\overline K=-i\epsilon_{\rm H}\int_0^1dt\,
   a^4\sum_x\lambda(x)\partial_\mu^*\overline k_\mu(x,t),
\eqn\sixteen
$$
because $\overline k_\mu$ is gauge invariant, i.e.,
$\delta_\lambda\overline k_\mu=0$. Since
$\partial_\mu^*\overline k_\mu(x)={\cal A}(x)$ from~Eq.~\fourteen\ for
anomaly-free cases, the gauge variation of $\overline K$~\sixteen\ in
fact cancels the gauge variation of~$W'$, Eq.~\ten. Namely, when the
anomaly cancellation condition {\it in the continuum field theory\/}
is fulfilled, the improved effective action~$W=W'+\overline K$ is
gauge invariant even with {\it finite\/} lattice spacing. The
improvement term~$\overline K$ does not spoil the desired properties
of~$W'$, because $\overline K$ contributes only the imaginary part
of~$W$ ($\overline k_\mu(x)$ transforms like the axial current), and
because the current~$\overline k_\mu(x)$ is higher order in~$a$, i.e.,
$\overline K\buildrel{a\to0}\over\rightarrow0$.

In the remainder of this paper, we show that the above expression of
the improved effective action~$W=W'+\overline K$ corresponds to the
formulation of~Ref.~[\LUSC], in the vacuum sector for an infinite
lattice volume. To see this, we consider a variation of the gauge
potential~$A_\mu(x)$, $\delta_\eta A_\mu(x)=\eta_\mu(x)$ and
$\delta_\eta A_\mu(x,t)=t\eta_\mu(x)$. The variation of the
functional~\six\ is then given by
$$
   \delta_\eta W'
   =\int_0^1dt\,\Tr\Bigl(
   \partial_t\delta_\eta{\cal D}{\cal P}_{\rm H}{\cal D}^{-1}
   -{\epsilon_{\rm H}\over2}a\partial_t{\cal D}\delta_\eta{\cal D}
   {\cal D}^{-1}
   -\partial_t{\cal D}{\cal P}_{\rm H}{\cal D}^{-1}
   \delta_\eta{\cal D}{\cal D}^{-1}\Bigr).
\eqn\seventeen
$$
The relation~\four\ follows
${\cal P}_{\rm H}{\cal D}^{-1}\delta_\eta{\cal D}=%
{\cal D}^{-1}\delta_\eta{\cal D}{\cal P}_{\rm H}-%
\epsilon_{\rm H}a\delta_\eta{\cal D}/2$. Therefore we have
$$
\eqalign{
   \delta_\eta W'
   &=\int_0^1dt\,(
   \partial_t\Tr\delta_\eta{\cal D}{\cal P}_{\rm H}{\cal D}^{-1}
   -\Tr\delta_\eta{\cal D}\partial_t{\cal P}_{\rm H}{\cal D}^{-1})
\cr
   &=\Tr\delta_\eta{\cal D}{\cal P}_{\rm H}{\cal D}^{-1}
   +\epsilon_{\rm H}\int_0^1dt\,{1\over2}a
   \Tr\delta_\eta{\cal D}\partial_t{\cal D}{\cal D}^{-1}.
\cr
}
\eqn\eighteen
$$
The second term here can be written from~Eq.~\four\ as
$$
\eqalign{
   {1\over2}a\Tr\delta_\eta{\cal D}\partial_t{\cal D}{\cal D}^{-1}
   &=-{1\over2}a\Tr\delta_\eta{\cal D}\partial_t{\cal D}
   {\mit\Gamma}_5{\cal D}^{-1}{\mit\Gamma}_5
   -{1\over2}a^2\Tr\delta_\eta{\cal D}\partial_t{\cal D}
   {\mit\Gamma}_5
\cr
   &=-{1\over4}a^2\Tr{\mit\Gamma}_5\delta_\eta{\cal D}
   \partial_t{\cal D}
\cr
   &=\epsilon_{\rm H}\Tr{\cal P}_{\rm H}
   [\partial_t{\cal P}_{\rm H},\delta_\eta{\cal P}_{\rm H}].
\cr
}
\eqn\nineteen
$$
Therefore the variation of~$W'$ is given by
$$
   \delta_\eta W'=\Tr\delta_\eta{\cal D}{\cal P}_{\rm H}{\cal D}^{-1}
   +\int_0^1dt\,\Tr{\cal P}_{\rm H}
   [\partial_t{\cal P}_{\rm H},\delta_\eta{\cal P}_{\rm H}].
\eqn\twenty
$$
Combined with the variation of~$\overline K$~\fifteen, the variation
of the total effective action may be expressed as
$$
\eqalign{
   \delta_\eta W&=\delta_\eta W'+\delta_\eta\overline K
\cr
   &\equiv\Tr\delta_\eta D{\cal P}_{\rm H}D^{-1}
   +i\epsilon_{\rm H}{\cal L}_\eta^\star,
\cr
}
\eqn\twentyone
$$
where
$$
\eqalign{
   &{\cal L}_\eta^\star
   \equiv a^4\sum_x\eta_\mu(x)j_\mu^\star(x)
\cr
   &=-i\epsilon_{\rm H}\int_0^1dt\,\Tr{\cal P}_{\rm H}
   [\partial_t{\cal P}_{\rm H},\delta_\eta{\cal P}_{\rm H}]
   +\int_0^1dt\,a^4\sum_x
   \bigl[\eta_\mu(x)\overline k_\mu(x,t)
   +A_\mu(x)\delta_\eta\overline k_\mu(x,t)\bigr]
\cr
   &=a^4\sum_x\eta_\mu(x)\overline k_\mu(x)
\cr
   &\quad-i\epsilon_{\rm H}\int_0^1dt\,\Tr{\cal P}_{\rm H}
   [\partial_t{\cal P}_{\rm H},\delta_\eta{\cal P}_{\rm H}]
   +\int_0^1dt\,a^4\sum_x
   \bigl[A_\mu(x)\delta_\eta\overline k_\mu(x,t)
   -\eta_\mu(x)t\partial_t\overline k_\mu(x,t)\bigr].
\cr
}
\eqn\twentytwo
$$
In deriving the last expression, we have performed a partial
integration by inserting~$1=\partial t/\partial t$. It is also easy to
see from Eq.~\eight\ that ${\cal L}_\eta^\star$ arises entirely\foot{%
For vector-like cases, the imaginary part of the effective
action~$W'$ vanishes identically, and thus $W'$ is gauge invariant.
Therefore ${\cal A}=\overline K=0$, and consequently
${\cal L}_\eta^\star=0$ in these cases~[\LUSC].} from~$\overline K$
and the imaginary part of~$W'$.

Equation~\twentytwo\ is identically the linear
functional~${\cal L}_\eta^\star$ in~Eq.~(5.8) of~Ref.~[\LUSC]. When
the lattice volume is infinite and the Dirac operator has no zero
modes, the variation of the effective action in~Ref.~[\LUSC] is given
by~Eq.~\twentyone. (See Eq.~(3.8) of~Ref.~[\LUSC].) Therefore, under
the above conditions, the effective action formulated by L\"uscher
can be represented as~$W=W'+\overline K$, i.e.,
Eq.~\six\ plus~Eq.~\fifteen. The content of~Theorem~5.3
of~Ref.~[\LUSC] also immediately follows in view of~Eq.~\twentyone:
(a)~When the gauge group is Abelian, the variation~$\delta_\eta$ and
the gauge variation~$\delta_\lambda$ commute if~$\eta$ does not
depend on the gauge potential. From this, we see that
$\delta_\eta W$~is gauge invariant because $W$~is gauge invariant. The
quantity $\Tr\delta_\eta D{\cal P}_{\rm H}D^{-1}$ is gauge covariant
by construction. This is equivalent to the gauge invariance in the
Abelian case. Therefore ${\cal L}_\eta^\star$~is gauge invariant.
(b)~${\cal L}_\eta^\star$ arises from the imaginary part of~$W$. Thus
it is consistent to assume $j_\mu^\star(x)$ (and
$\overline k_\mu(x)$) transforms like the axial vector current.
(c)~$(\delta_\eta\delta_\zeta-\delta_\zeta\delta_\eta)W=0$ for the
$A_\mu$-independent parameters~$\eta$ and~$\zeta$. Using a
calculation that is almost the same as that producing~Eq.~\nineteen\
{}from Eq.~\seventeen\ immediately shows
$$
   \delta_\eta\Tr\delta_\zeta D{\cal P}_{\rm H}D^{-1}
   -\delta_\zeta\Tr\delta_\eta D{\cal P}_{\rm H}D^{-1}
   =-\Tr{\cal P}_{\rm H}
   [\delta_\eta{\cal P}_{\rm H},\delta_\zeta{\cal P}_{\rm H}].
\eqn\twentythree
$$
Therefore ${\cal L}_\eta^\star$~satisfies the integrability condition
Eq.~(5.9) of~Ref.~[\LUSC]. (d)~The anomalous conservation
law~$\partial_\mu^*j_\mu^\star(x)={\cal A}(x)$ holds (when
$\sum_\alpha e_\alpha^3=0$) because $W$~is gauge invariant, and the
first term of~Eq.~\twentyone\ produces the gauge anomaly under the
gauge variation $\delta_\lambda D=-i[\lambda T,D]$, as in~Eq.~\ten.

The last line of~Eq.~\twentytwo\ implies a difference between the
``covariant gauge current,''
$\Tr\delta_\eta D{\cal P}_{\rm H}D^{-1}%
+i\epsilon_{\rm H}a^4\sum_x\eta_\mu\overline k_\mu$, and the
``consistent gauge current,''
$\Tr\delta_\eta D{\cal P}_{\rm H}D^{-1}%
+i\epsilon_{\rm H}a^4\sum_x\eta_\mu j_\mu^\star$. This quantity is
analogous to the quantity in the continuum field theory that relates
the covariant anomaly and the consistent
anomaly~[\BAR--\LEU,\BAN]. Interestingly, the difference does
not contribute to the integral~\fifteen, as can easily be
verified~[\BAN]. Thus we may use $j_\mu^\star(x)$ instead
of~$\overline k_\mu(x)$ in~Eq.~\fifteen. Of course, the structure
of~$\overline k_\mu(x)$ is simpler and the expression of
$j_\mu^\star(x)$~\twentytwo\ is not needed in our
representation~$W=W'+\overline K$. The existence of the ``integrable
current'' $j_\mu^\star(x)$ is important in ensuring~[\LUSC] the
existence of the path integral expression that corresponds to the
effective action~$W$.

It is certainly desirable to perform the present analysis in finite
volume.\foot{%
Some steps of our discussion may be formal due to possible IR
divergences. This also prompts us to pursue an analysis on a finite
volume lattice.} The representation~\nine\ might be useful in
identifying a possible lattice counterpart of the $\eta$-invariant.
Obtaining a lattice implementation of the $t$-integrals in~Eqs.~\six\
and~\fifteen\ is also an interesting problem. We postpone these for
future projects.

The author is grateful to Dr.~Y.~Kikukawa for discussions and to
Professor~K.~Fujikawa for encouragement. This work is supported in
part by the Ministry of Education Grant-in-Aid for Scientific
Research~Nos.~09740187 and~10120201.

\refout
\bye